\documentclass{aastex}
\usepackage{emulateapj5,epsfig}
\usepackage{apjfonts}
 
\begin{document}
\makeatletter
\newenvironment{inlinetable}{%
\def\@captype{table}%
\noindent\begin{minipage}{0.999\linewidth}\begin{center}\footnotesize}
{\end{center}\end{minipage}\smallskip}

\newenvironment{inlinefigure}{%
\def\@captype{figure}%
\noindent\begin{minipage}{0.999\linewidth}\begin{center}}
{\end{center}\end{minipage}\smallskip}
\makeatother

\def\wdf{white dwarf}
\def\etal{et al.}
\def\rd{Di\thinspace Stefano}

\title{Supersoft and Quasisoft X-Ray Sources in the Globular Clusters of  
NGC 4472:
Are they candidates for intermediate-mass black holes?}

\author{R.~Di\,Stefano\altaffilmark{1,2}, R. Friedman\altaffilmark{3}, A. Kundu \altaffilmark{4} and A.~K.~H.~Kong\altaffilmark{1}}

\altaffiltext{1}{Harvard-Smithsonian Center for Astrophysics, 60
Garden Street, Cambridge, MA 02138}
\altaffiltext{2}{Department of Physics and Astronomy, Tufts University, Medford, MA 02155}
\altaffiltext{3}{Department of Astronomy and Astrophysics, University of Chicago, 
5640 S. Ellis Ave, Chicago, IL 60637}
\altaffiltext{4}{Department of Physics and Astronomy, 
Michigan State University, East Lansing, MI 48824}


\begin{abstract}
	We report on possible associations between $6$ globular clusters
in the Virgo elliptical galaxy NGC 4472 (M49) and bright ($L_x > 10^{38}$
erg s$^{-1}$) very soft X-ray sources (VSSs).
Two of the VSSs have broad-band spectral properties
consistent with those of 
luminous
supersoft X-ray sources (SSSs). 
The other VSSs are somewhat harder harder, possibly with
values of $k\, T$ between roughly $150$ eV and $250$ eV.
These sources may be too hot to be explained
by the white dwarf models so promising for SSSs; they  are members of 
the newly-established class of quasisoft sources (QSSs).  
We examine white dwarf, neutron star, and
black hole  models for the VSSs. One of the SSSs is hot and bright enough
to be a possible progenitor of a Type Ia supernova, while the most natural model
for the other VSSs is one in which the accretors are intermediate-mass 
black holes.        
Whatever their physical natures, these sources are unlike any X-ray 
sources in Galactic globular clusters. No Galactic globular
cluster houses a quasisoft
source and only one, M3, contains
a dim 
($L_x \sim 10^{36}$
erg s$^{-1}$) transient SSS.
\end{abstract}


\section{Introduction}

\subsection{Very Soft X-Ray Sources}

Very soft X-ray sources (VSSs) were observed by the {\it Einstein X-ray Observatory}
in the Magellanic Clouds (MCs; Long, Helfand, \& Grabelsky 1981;
Seward \& Mitchell 1981).
When {\it ROSAT's} {\it All Sky Survey}
discovered roughly $30$ such soft X-ray,
sources in the 
MCs, M31, and in the Galaxy,    
the class of luminous supersoft X-ray sources (SSSs) was established. 
(See Tr\"umper et al.\, 1991,
Supper et al\, 1997, and references in Greiner 2000.)
The luminosities
of the first SSSs to be discovered were in the range from
$10^{37}$ ergs s$^{-1}$ to a few times $10^{38}$ ergs s$^{-1}$, with
$k\, T$ in the range of tens of eV.
With the advent of {\it Chandra} and {\it XMM-Newton,} 
the study of SSSs can be extended to
more distant galaxies.
We have therefore developed a systematic method to select soft sources
from the pool of all X-ray sources detected by either
{\it Chandra} or {\it XMM-Newton} in an external galaxy
(\rd\ \& Kong 2003 a, b, c).
We have applied this algorithm
to M31 (\rd\ et al.\ 2003 a); 
M104 (\rd\ et al.\ 2003 b); NGC 300 (Kong \& \rd\ 2003);  
M101, M83, M51, NGC 4697 (\rd\ \& Kong 2003 c); 
and NGC 4472 (Friedman et al.\, 2002).
We find that the algorithm 
selects not only SSSs, but also members of a new class
of sources which have been dubbed {\it quasisoft sources} (QSSs). QSSs 
are luminous ($L$ typically $> 10^{36}$ erg s$^{-1}$)
sources that
appear to be slightly harder ($150-250$ eV) than SSSs, yet soft enough that
their emission falls off at energies well  below the
energies characteristic of canonical X-ray binaries.    
We use the term VSS for any source that is either supersoft or quasisoft.

\subsubsection{Models for Supersoft Sources} 

The measured values of $L$ and $T$ 	
suggest
white dwarf (WD) models.  
Indeed, almost half of the local SSSs
with optical IDs are systems housing a hot WD or pre-WD (e.g.,
recent novae, symbiotics, planetary nebulae). (See Greiner's [2000] SSS 
Catalog.)  Perhaps the most
promising model for the remaining  
$9$ SSSs with optical IDs is one in which a
Roche-lobe-filling companion donates
mass to a WD at rates high enough 
to allow quasi-steady nuclear burning (van den Heuvel et al.\, 1992;
Rappaport \rd\ \& Smith 1994).  
Matter which
is burned can increase the mass of the WD, potentially leading to a Type Ia
supernova explosion. Nevertheless, 
neutron star (NS) models are not excluded
(Kylafis \& Xilouris 1993), and there have been suggestions
that some of the local SSSs may be accreting stellar-mass black holes 
(BHs; Cowley, Schmidtke,
 Crampton, \& Hutchings 1990, 1998; Hutchings, Crampton,
 Cowley, \& Schmidtke 1998). 

In external galaxies
we can study the positions of SSSs relative to 
other galaxy inhabitants. In M101, M83, and M51, 
some SSSs 
are located in the spiral arms, some 
close to markers of young stellar populations. While some of these
may be supernova remnants, the fact that many SSSs 
in external
galaxies are variable suggests 
that are X-ray binaries.
(Variability studies have been possible so far for SSSs in M31
and for the high-$L$ sources in a handful of nearby galaxies.)
At least some accretors may therefore be NSs or BHs in young ($< 10^8$ yrs).
In addition, some SSSs are ultraluminous, suggesting accretion onto an
intermediate mass ($100-1000\, M_\odot$) BH. The ultraluminous SSSs
represent an extension of the SSS class to luminosities so high that
the accretors are almost certainly not WDs. In an analogous way, QSSs
represent an extension to higher energy emission that also appears
to be inconsistent with WD models. As discussed below (see Eq.\ 1),
both the extension to higher luminosities and the extension to higher
temperatures are consistent with models of accreting
intermediate-mass BHs (IMBHs).    

There are  SSSs positioned  within a few parsecs of the central
black holes of some galaxies (e.g., M31 [Garcia et al. 2000; Kong et al. 2002]); 
these may be the stripped cores of tidally disrupted giants (\rd\ et al. 2001).  

\subsubsection{Models for Quasisoft Sources}

The luminosity and temperature of nuclear burning WDs both
increase with the WD mass. Since the WD mass cannot be
larger than 
$1.4\, M_\odot,$ the Eddington limit places an upper bound on the
luminosity of $2-3 \times 10^{38}$ erg s$^{-1}$. Because
the photospheric radius has a lower bound equal to the WD radius,
this places an upper bound on the effective temperature of approximately
$150$ eV. In fact, this temperature is
not likely to be achieved by a nuclear-burning WD, since the
photosphere is generally above the WD's surface.
Sources
with higher $T$ 
and effective radii too small to be consistent with WDs,
are likely to be accreting NSs or BHs.
NS or stellar-mass BH accretors 
do not seem natural, because the 
photospheric radii would be $2 - 3$ orders of magnitude
larger than the NS or Schwarzschild radii.
There is, however, one natural model: if the accretor 
is an IMBH, then the temperature and luminosities 
are expected to be in the range observed for both SSSs and QSSs.
Let $M_{BH}$ be the mass of the accreting BH, and let $\alpha$ be the 
efficiency, with $L=\alpha \dot M_{BH}\, c^2.$ If we assume that the
accretion is mediated by a thin disk, and that the disk is optically thick,
with its inner edge coincident with the radius of the last stable orbit,
then the temperature of the inner disk can be
written as follows.
\begin{equation} 
k\, T > 158\, eV \Bigg({{100 M_\odot}\over{M_{BH}}}\Bigg)^2\, 
\Bigg({{L}\over{3.1 \times 10^{37} erg s^{-1}}}\Bigg)^{{1}\over{2}}
\Bigg({{0.1}\over{\alpha}}\Bigg)^{{1}\over{2}}
\end{equation} 
A better model is a multi-color disk, generally with a power law tail 
(Mitsuda et al. 1984).
Using a more sophisticated model than the one that produced Eq.\ 1,
by, e.g., taking effects such as spectral hardening or the possible BH
spin into account, would tend to increase the  
estimated BH mass for any set of measured values of $k\, T$ or $L$; 
this is the reason for the inequality.
The higher the mass of the BH, the lower the temperature.

\subsection{SSSs in Galactic GCs} 

\subsubsection{Observations}

Only one SSS has been detected in a Galactic globular cluster (GC). 
1E 1339.8+2837,
located in M3 (NGC 5272), was observed both in
the ROSAT all-sky survey (Verbunt et al. 1995), and in
pointed observations using the HRI (Hertz, Grindlay \& Bailyn 1993).
Estimates of the temperature and bolometric luminosity based
on the all-sky survey were $k T \sim 45$ eV, and $L \sim 10^{35}$
erg s$^{-1}$ (Verbunt et al. 1995);                                                                   
 estimates based on the HRI observations
were $k T \sim 20$ eV and $L \sim 1.6 \times 10^{36}$ erg s$^{-1}$ (Hertz et al. 1993).
The source is a transient.
Note that the inferred luminosity (though uncertain)
is smaller than typical for SSSs
found in the Galaxy, the MCs , and M31.
This source could be similar to V751 Cyg, a nova-like variable of the
 VY Scl class, which becomes a low-$L$ SSS only during rare
optical low states (Greiner et al. 1999, Greiner \& \rd\ 1999).   

Because GCs have been well-studied at X-ray wavelengths, we can 
determine whether systems like 1E 1339.8+2837 are common.
By carrying out simulations in which GCs were
``seeded'' with SSSs (see \rd\ \& Rappaport 1994 for details),
\rd\ \& Davies (1996) concluded that {\it ROSAT} observations
of GCs 
should have produced an almost complete census
of any active GC SSSs with characteristics similar to
either 1E 1339.8+2837 or to the other known SSSs.
Thus, SSSs in Galactic GCs are rare.

\subsubsection{Theory}

Globular clusters have per capita populations 
of low-mass X-ray binaries (LMXBs)
roughly $100$ times larger than in the Galactic
disk. The enhancement is due to 
stellar interactions
within or near the dense cluster cores.
If the same ratio held for SSSs, then there would
be $\sim 100$ SSSs in Galactic GCs. With just one 
observed, 
we can rule out the possibility that the 
stellar interactions are producing significant enhancements
in the numbers of GC SSSs.

It is  nevertheless worth noting that SSSs {\it can} be produced by $3$ or
$4$ body interactions in GCs. 
The donor star would be a blue straggler,
formed through the merging of two lower-mass stars.
(In fact, M3
is rich in blue stragglers [Guhathakurta {et al.} 1994].)
Di\thinspace Stefano \&
Davies (1996) carried out simulations
of the formation and binary evolution of GC systems that could
become SSSs. 
They found that there should ne
 $\sim 1$ SSS in Galactic GCs, 
comparable to predictions of the numbers of SSSs expected due to
ordinary binary evolution, and consistent with the X-ray observations. 
In addition, GC SSSs were likely 
to have values of $\dot m$
below the steady-burning region, and should therefore appear as SSS transients.

\section{NGC 4472}
\subsection{NGC 4472 as a Testing Ground}
NGC 4472 is a giant elliptical galaxy in the Virgo cluster. 
It houses
about 6000 GCs (Rhode \& Zepf 2001). {\it Chandra}
discovered 144 X-ray sources in
an $8'\times8'$ region near the center of the galaxy
(Kundu et al. 2002; Maccarone, Kundu, \& Zepf 2003). There are more than 
800 GCs in the region that {\it Chandra} observed (Kundu \& Whitmore 2001).
In the regions studied by both {\it Chandra} and {\it HST},
40\% of the bright
($> 10^{37}$
erg s$^{-1}$) X-ray sources are associated with optically identified GCs
(Kundu et al. 2002; Maccarone, Kundu, \& Zepf 2003).

We conducted an independent analysis with two new elements.
First, we considered data from all $6$ CCDs.
Second, we 
ran the detection algorithm (WAVDETECT) for energies 
in the range $0.1-7$ keV,
rather than $0.5-7$ keV. 
Because emission from the softest sources
begins to fall off by $0.5$ keV, photons
below $0.5$ keV may combine with photons of higher 
energies, to make detections possible.
To ensure that none of the sources are artifacts,
we visually inspected the 
region containing each source, to verify that the spatial distribution
of photons
was consistent with the PSF at the source location.

We used lists of HST-identified GCs from
Kundu et al. (2002), and of spectroscopically studied GCs
from Sharples et al. (1998) and Zepf et al. (2000).
To conduct a search for GCs in the
vicinity of each X-ray source, we considered a sequence
of search cones of radius $r$ ($0.5'' < r < 5''$). The optimal value of $r$
was $1.15''.$ Using this value we found $40$ GCs near
X-ray sources. An analysis using random shifts
of the X-ray source positions, found that that $\sim 6$
identifications are likely to be due to chance superpositions.

\subsection{VSSs in NGC 4472}
We applied a uniform selection procedure
(Di\,Stefano \& Kong 2003a, b)
 to determine which
X-ray sources are VSSs.
Our selection algorithm identified $27$ SSSs in NGC 4472. Six
of these are associated with GCs.
Some of the VSS-GC pairs may be artifacts. Since,
however, we have estimated the number of
spurious pairings to be $6$,
it is unlikely that 
all $6$ false matches involve VSSs. We therefore conclude that
at least several of the VSS-GC matches represent genuine
physical correspondences.
Table 1 lists the SSSs identified with NGC 4472's GCs.
Sources $4$ and $5$ were identified as SSSs; the remaining sources
were identified as QSSs. 

Some of the VSSs may be background active galactic nuclei.
We used results from the Chandra Deep Field Surveys (e.g.,
Brandt et al. 2001; Giacconi et al. 2001) to find 
that, at $4.5\times10^{37}$ erg s$^{-1}$ (the completeness limit),
fewer than 10 sources are likely to be background objects.
We expect that only a small fraction of these 
are soft enough to be identified as VSSs.
To quantify this, we studied data from several
fields observed by {\it Chandra}
\footnote{http://hea-www.cfa.harvard.edu/CHAMP}
and find that $1-3$ VSSs unrelated to NGC 4472 are likely to be
present in our data; these are most likely to be foreground stars (see
e.g. Di\,Stefano et al. 2003a and Kong et al. 2003). 

We have also investigated the optical properties of the 
GCs that house VSSs.
We find that the VSSs are preferentially in luminous clusters.
This parallels results for X-ray sources in general. (See e.g.,
\rd\ et al.\ 2002.) The VSSs are found in both metal-poor
and metal-rich GCs.

\section{Models and Implications}
Because the portion of the galaxy for which we have GC positions
is just a small portion of the region covered by the {\it Chandra}
observations, any estimate of the fraction of X-ray sources
residing in GCs is highly uncertain. 
Note however, that roughly $1/5$ of the SSSs 
and $1/6$ of all X-ray sources are
associated with GCs.
The fact that these $2$ ratios are similar, and that, as in
our Galaxy, the 
X-ray sources are over-represented in GCs of by a factor   
of $\sim 100,$ suggests that X-ray sources in general,
and SSSs in particular, are produced in the GCs of NGC 4472 through 
interactions. This is different from the situation in
the Milky Way, where X-ray sources in general, but not SSSs, are
overproduced in GCs.

\subsection{Broad-Band Spectra}

The number of counts from each GC SSS is too small to
permit a spectral fit. We can, however, compare the distribution
of counts from each source in the S (0.1-1.1 keV), M (1.1-2 keV), 
and H (2-7 keV) bands,
with the distributions expected from known models. 
To this end, we used PIMMS (AO3 release)
to compute the numbers of counts expected in S, M, and H from
a 
blackbody source in NGC 4472, located 
behind a column with $N_H$ equal to each of $4$ values:
($4\times10^{20}$ cm$^{-2}$, $1.6\times10^{21}$ cm$^{-2}$, $6.4\times10^{21}$ cm$^{-2}$ 
and  $2.5\times10^{22}$ cm$^{-2}$). 
The values of $k\, T$ we considered 
ranged from 25 eV 
to 100 eV in 15 eV increments, then to 250 eV is $25$ eV increments, and then
to 500 eV and beyond by increments of 250eV. 
For each VSS,
we scaled the luminosity of the model source so that
it produced the same total number of counts.
If the computed counts in each band were within $1-\sigma$ 
of the observed counts, 
we counted the model as a possible match.
The results are shown in Figure 1. 

\subsection{Physical Models} 

\subsubsection{Supersoft Sources}

The softest source is source 5, with the associated models 
shown as magenta pentagons. The
fits 
are highly degenerate. Source $5$ could be either an ultraluminous source with
a very low value of $k\, T$, or a source with the characteristics
of a classical SSS.  
In the latter case, it is likely to 
have $k\, T$ near or just under $100$ eV, and to have $L$ close to
the Eddington limit for a $1.4\, M_\odot$ object. 
Thus, this particular SSS could be an accreting high-mass WD, perhaps
a good candidate for the progenitor af a Type Ia supernova explosion.
It has not been anticipated that GCs serve as sites of 
Type Ia SNe. Nevertheless, we can construct scenarios in which progenitors
can be formed in GCs. If the SSS is WD binary with orbital period
less than a few days, it is likely that the donor is a $1-2\, M_\odot$
star that may be slightly evolved. This is possible if the cluster
is young, or if the donor is a blue straggler (see \S 1.2.2).
If, instead, source $5$ is very soft (with $k\, T$ possibly as low as $10$ eV),
and very luminous, then the most likely interpretation is that it is
an IMBH, 
with mass that could be as high as $10^3-10^4 M_\odot.$ 
Although the presence of IMBHs in GCs has been suggested,
the more conservative model for this source is likely to be the WD model.

Source $4$ 
(cyan boxes)  is also chosen as an SSS.
It appears to be  not as soft as source $5$;
values of $k\, T < 100$ eV are ruled out. This means that
the range of possible luminosities is also
more restricted. If source $4$ is a nuclear-burning WD it
is marginally hotter, while at the same time
being somewhat less luminous, than expected for a high-mass
WD. In fact, source $4$ is consistent with
models very similar to those that apply to the QSSs, discussed below. 

\vspace{0.6cm}
\begin{inlinefigure}
\psfig{file=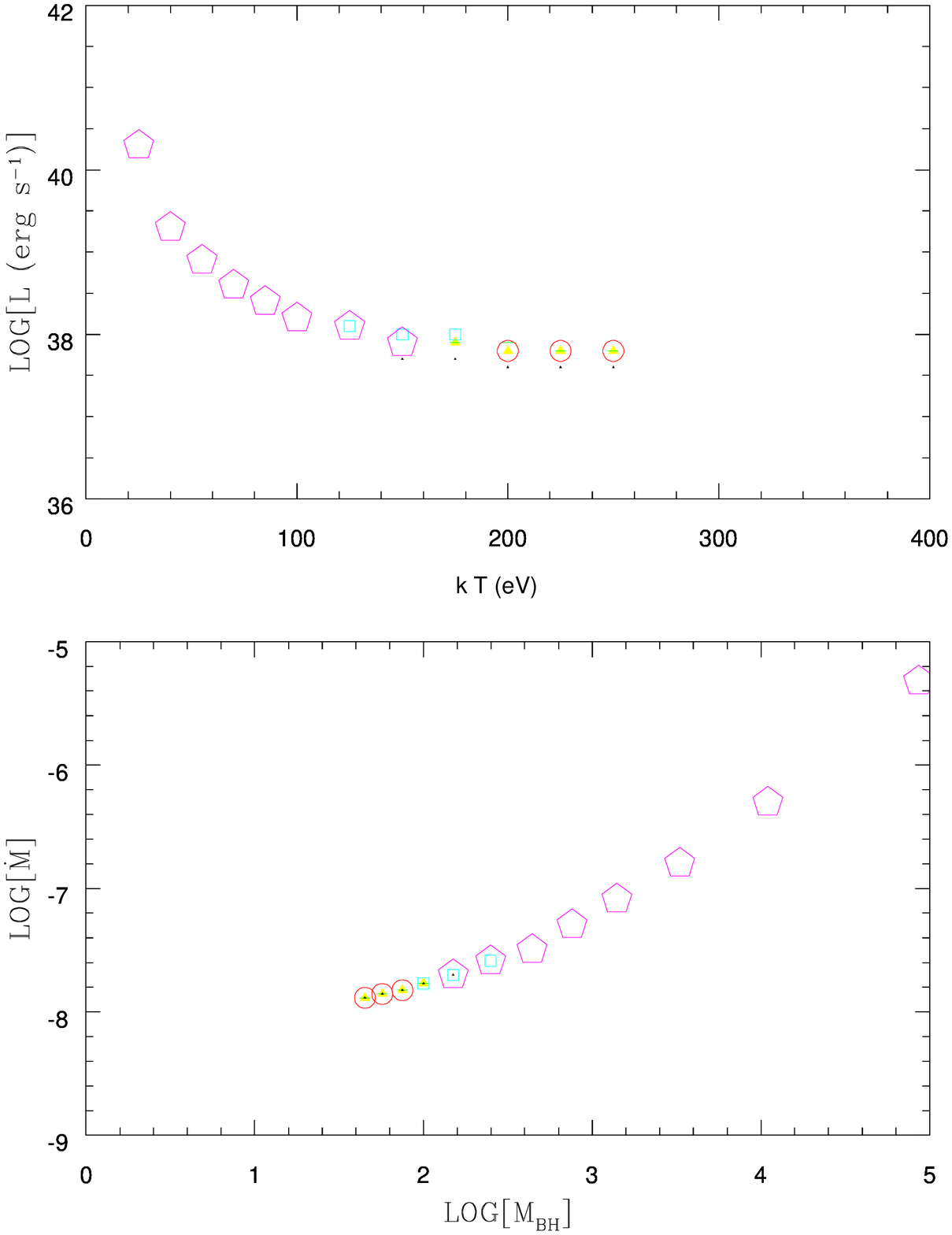,height=2.8in}
\vspace{1.3cm}
\caption{
Red circles show models that match source 1; yellow triangles, 2; green lines, 3; 
cyan boxes, 4; magenta pentagons, 5;
small black triangles, 6. The top panel shows the results of matching thermal
models to the broad-band data. Shown for each source are the models that
produced count rates in each band (S, M, and H), within
$1-\sigma$ of the observed counts. Log[L] is  plotted vs. $k\, T.$      
Each source is compatible with more than one model. In the bottom
panel, Log[$\dot M$] is plotted against the log of the minimum
BH mass;
each point corresponds to a model shown in the top panel. Equation (1)
was used to compute the minimum
BH mass. To compute $\dot M,$ we assumed that $L= 0.1\, \dot M\, c^2.$ 
}
\end{inlinefigure}

\subsubsection{Quasisoft Sources}

The QSSs appear to be too hot to be compatible with
WD models. The associated values of $k\, T$ appear to be clustered between
roughly $150$ eV and $250$ eV, while the luminosities
all lie near $10^{38}$ erg s$^{-1}$. Although
the nature of these sources is
not determined,  
IMBH models are well suited to them. The minimum computed values of the BH masses
range from approximately $50\, M_\odot$ to $150\, M_\odot.$
The mass accretion rates would be similar among the systems, around
$10^{-8} M_\odot$ yr$^{-1}$.

\section{Conclusions} 

One VSS has already been associated with a GC in the 
Sombrero galaxy, M104 (\rd\ et al.\, 2003). Nevertheless, 
the discovery of a set of VSSs in the GCs of NGC 4472 is remarkable.
First, the discovery of $6$ VSS-GC matches rules out the
possibility that all of them are spurious. In combination
with the M104 results, we can now more firmly 
establish that GCs do house luminous 
($\sim 10^{38}$ erg s$^{-1}$) VSSs. There is no precedent for
this in the Galaxy or in M31.  
Second, our NGC 4472 results provide examples of VSS-GC matches
in an elliptical galaxy. 
Third, the set of GC-VSSs allows us to begin studying the
statistics of this connection. Much more data, from a larger
number of galaxies is needed. But, e.g., if it remains the case
that there is no obvious connection between GC metalicity and
the likelihood of hosting an SSS, this would be a significant
contrast with other X-ray binaries, suggesting
that different processes create the VSSs than those that create other
X-ray binaries. 

If some of the VSSs in GCs
are WDs accreting matter at high enough rates to allow
nuclear burning, then Type Ia SNe may occur in GCs.
Young GCs, or extremely massive centrally dense GCs would be favored.
If Type Ia supernovae can occur in GCs, it is possible that 
surveys designed to study Type Ia supernovae will  discover an example.  

Most sources in the 
range $150-250$ eV sources 
are likely not hot WDs. The most
natural alternatives are accreting IMBHs.
It is interesting that the properties of the QSSs in GCs
seem roughly similar across clusters, and that the
minimum values of the BH masses are also similar ($50-150\, M_\odot$). 

The presence of IMBHs in GCs has been predicted by a number
of authors. Dynamical evidence for such BHs in M15 and
G1 has been offered (Gerssen et al. 2002; Gebhardt, Rich, \& Ho 2002), 
although these results have 
generated some
controversy (McNamara, Harrison, \& Anderson 2003; Baumgardt et 
al. 2003). The discovery of 
SSSs, and especially of QSSs, 
in GCs may provide independent observational evidence that
BHs with masses larger than standard stellar remnants do inhabit
GCs. If so, studies like the one described here
will provide an avenue to establish the statistics and 
general properties of the 
connection between globular clusters and intermediate-mass
black holes.

\begin{acknowledgements}
It is a pleasure to acknowledge conversations with P. Barmby, 
which helped us to select NGC 4472 as a target of study.  
This work was supported by NASA under an LTSA grant,
NAG5-10705 and by NSF grant AST-9731923 to the SAO Summer
Intern program. A.K. acknowledges NASA for support via LTSA grant NAG5-12795. 
A.K.H.K. acknowledges support from the Croucher Foundation.
\end{acknowledgements}

\begin{deluxetable}{cccccccc}
\tabletypesize{\small}
\tablecaption{VSSs in NGC 4472's GCs}
\tablewidth{0pt}
\tablehead{\multicolumn{1}{c}{Source}& R.A.& Dec. & \multicolumn{3}{c}{Net 
Counts} & \multicolumn{2}{c}{Hardness Ratios}\\
& (h:m:s)& $(^{\circ}:\arcmin:\arcsec)$ & Soft & Medium & Hard & HR1 & 
HR2}
\startdata
1 & 12:29:47.8& +07:59:19& $9.26\pm2.43$ & $5.84\pm1.91$ & $1.00\pm0.40$ & 
-0.23 & -0.81\\
2 & 12:29:47.8& +07:59:44& $9.27\pm2.62$& $5.55\pm1.87$ & $0.78\pm0.31$ & 
-0.25 & -0.84\\
3 &12:29:40.1 & +07:58:29& $12.91\pm2.97$ & $5.46\pm1.82$ & $0.00\pm0.00$ 
& -0.41 & -1.00 \\
4 &12:29:48.6 & +07:59:13& $15.52\pm3.36$ & $2.62\pm1.10$ & $1.28\pm0.53$ 
& -0.71 & -0.85\\
5 &12:29:45.0 & +08:01:06& $15.26\pm3.25$ & $1.09\pm0.44$ & $0.00\pm0.00$ 
& -0.87 & -1.00\\
6 & 12:29:55.8& +07:57:23& $6.46\pm1.89$ & $3.28\pm1.24$ & $0.00\pm0.00$ & 
-0.33 & -1.00\\
\enddata
\normalsize
\tablecomments{
The
columns give the source number, the position (J2000.0), the net counts
in the three energy bands (soft: 0.1--1.1 keV; medium: 1.1--2 keV;  hard: 
2--7 keV), and the hardness ratios (HR1: (medium-soft)/(medium+soft); HR2: 
(hard-soft)/(hard+soft). Note that the selection procedure does not rely solely on hardness ratios (see \rd\ \& Kong 2003a,b).}
\end{deluxetable}

\end{document}